\documentclass{aastex}
\singlespace
\slugcomment{Accepted for the publication in the {\it Astrophysical Journal}}
\newcommand{\hii}{H{\small II} }

\def\farcs{\hbox{$.\mkern-4mu^{\prime\prime}$}}

\def\la{\mathrel{\hbox{\rlap{\hbox{\lower4pt\hbox{$\sim$}}}\hbox{$<$}}}}
\def\ga{\mathrel{\hbox{\rlap{\hbox{\lower4pt\hbox{$\sim$}}}\hbox{$>$}}}}

\shortauthors{Park}
\shorttitle{SN 1987A}

\begin{document}
\title{Monitoring the Evolution of the X-ray Remnant of SN 1987A}
\author{Sangwook Park, David N. Burrows, Gordon P. Garmire,
and John A. Nousek}
\affil{Department of Astronomy and Astrophysics, Pennsylvania State
University, 525 Davey Laboratory, University Park, PA. 16802}


\author{and}

\author{Richard McCray, Eli Michael, and Svetozar Zhekov}

\affil{Joint Institute for Laboratory Astrophysics, University of
Colorado, Campus Box 440, Boulder, CO. 80309-0440}

\begin{abstract}

We report on the results of our monitoring program of the remnant
of SN 1987A with the Advanced CCD Imaging Spectrometer (ACIS)
on board the {\it Chandra X-ray Observatory}. Two new observations
have been performed in AO2, bringing the total to four monitoring
observations over the past two years. Over this time period, new 
techniques for correction of ``Charge Transfer Inefficiency (CTI)'' 
and for use of charge spreading to provide angular resolution 
somewhat better than the pixel size of the CCD detector have 
become available at Penn State. 
We have processed all four observations using sub-pixel
resolution to obtain the highest possible angular resolution, and
using our CTI correction software to provide more reliable spectral
analysis and flux estimations.

The high angular resolution images indicate that the X-ray bright knots
are convincingly correlated with the optical spots, primarily at
$\la$1 keV, while higher energy photons are very well correlated with
radio images. Our data also provide marginal evidence for radial 
expansion of the X-ray remnant at a rate of 5200 $\pm$ 2100 km s$^{-1}$.
The X-ray flux appears to linearly increase by $\sim$60\% over the 
18 month period of these observations. 
The spectrum is dominated by broad complexes of
atomic emission lines and can be fit with a simple
model of a plane-parallel shock with electron temperatures of $kT$ 
$\sim$ 2 $-$ 4 keV and a postshock electron density of $n_{e}$ $\sim$ 210
$-$ 420 cm$^{-3}$. The implied 0.5 $-$ 10 keV band luminosity 
in 2001 April is $\sim$1.3 $\times$ 10$^{35}$ ergs s$^{-1}$; as of
that date, we still observe no direct evidence for the central point source,
with an upper limit on the {\it observed} luminosity of $L_{X}$ 
$\sim$ 5.5 $\times$ 10$^{33}$ ergs s$^{-1}$ in the 2 $-$ 10 keV band.

\end{abstract}


\section {\label {sec:intro} INTRODUCTION}

With a known age (the supernova explosion in 1987; Shelton et al. 1987),
distance ($\sim$50 kpc in the Large Magellanic Cloud [LMC]; Andreani et
al. 1987), and progenitor (Sanduleak $-$69$^{\circ}$202, a type B3 I star)
\citep{kirshner87,sonneborn87}, SN 1987A provides a very rare yet
excellent astrophysical laboratory for the study of the evolution of
a supernova remnant (SNR) in the early phase. The supernova explosion
for SN 1987A has also confirmed core collapse in Type II SN
by the detection of a neutrino
burst \citep{koshiba87}. Since its discovery, SN 1987A has thus been
intensively studied with space-borne and ground-based instruments over
the entire electromagnetic spectrum (e.g., Chevalier 1992 and
references therein).

Optical observations with the {\it Hubble Space Telescope} (HST)
have shown an elliptical inner ring around the remnant as well as two
closed outer loops \citep{burrows95}. The inner ring is believed
to be a result of the interactions of the stellar winds from two
phases of the progenitor \citep{luo91,wang92,blondin93}. 
Inside of this ring is an \hii region produced by the UV radiation
from the progenitor star \citep{chevalier95}, with rapidly expanding 
ejecta clearly visible at the center of {\it HST} images.
The broad high-velocity (4000 km s$^{-1}$ $-$
15000 km s$^{-1}$) Ly$\alpha$ and H$\alpha$ emission detected
with the {\it HST} Space Telescope Imaging Spectrograph (STIS)
comes from a reverse shock located at $\sim$75\% of the
inner ring radius \citep{michael98}, which is propagating into
the ejecta. The {\it HST} data have also exhibited the emergence 
of several ``optical spots'' along the inner ring since the 
first detection of such an optically bright spot in 1997 
\citep{pun97,garnavich97}.
These optically brightening spots are interpreted as the emission by
radiative shocks as the supernova blast wave begins to strike inward
protrusions of the dense material ($n$ $\sim$ 10$^{4}$ cm$^{-3}$)
in the inner ring \citep{michael00a, michael00b, pun01}. 
The continuing developments of the optical spots \citep{lawrence00,
bouchet00, maran00,garnavich00} thus signal the precursor of
an exciting event: a predicted dramatic brightening (up to the 3 orders of 
magnitudes) of the remnant at almost all wavelengths \citep{luo94,
borkowski97a} in the near future as the shock front hits the ring proper.

The X-ray emission from SN 1987A was detected with {\it ROSAT}
and the X-ray lightcurve revealed a steady increase of the
soft X-ray flux over the 4-year {\it ROSAT} observation period
\citep{hasinger96}. The {\it ROSAT} observations, however, were unable
to resolve the X-ray remnant due to its limited angular resolution
($\sim 5 ''$), or to determine the nature of the emission due to the
low spectral resolution of the PSPC \citep{hasinger96}.

Recently, unprecedented high-angular resolution images from
observations of SN 1987A with the {\it Chandra X-ray Observatory}
have revealed a shell-like structure of the X-ray remnant which appears
to peak just inside of the optical inner ring \citep{burrows00} (B00
hereafter). The size of the shell structure is 1$\farcs$2 $\times$
1$\farcs$0 and the X-ray bright ``knots'' are roughly associated with 
the optical spots within uncertainties of $\sim$0$\farcs$1. 
The 0.5 $-$ 2 keV X-ray lightcurve is now increasing much more rapidly 
than the linear extrapolation of the {\it ROSAT} lightcurve.
The high-resolution dispersed spectrum is dominated by highly ionized 
atomic line emissions (B00), indicating a thermal origin of the X-ray 
emission. The implied electron temperature was $kT$ $\sim$ 3 keV. 

In 2000 December and 2001 April, two new {\it Chandra} observations
of SN 1987A have been performed for the purpose of monitoring the
development of the X-ray remnant and we here report the results of the
image and the preliminary spectral analysis.
The observations are described in \S\ref{sec:obs}.
The analysis and results are presented in \S\ref{sec:analysis} and the
implications are discussed in \S\ref{sec:disc}. A summary and the
conclusions are presented in \S\ref{sec:summary}.

\section{\label{sec:obs} OBSERVATIONS \& DATA REDUCTION}

As part of a monitoring program of SN 1987A, we have
performed a total of four observations with the Advanced CCD Imaging
Spectrometer (ACIS) on board the {\it Chandra
X-ray Observatory} \citep{weisskopf96} between 1999 and 2001 (Table
\ref{tbl:tab1}).
The first two observations were carried out in 1999 October and 2000
January as a part of the {\it Chandra} Guaranteed Time Observation (GTO)
program. The first observation was taken using the High-Energy
Transmission Grating (HETG) and the ACIS-S detector array. The second
observation used the ACIS-S3 detector without grating.
The detailed description of the observations and some preliminary
results have been presented in B00. Two subsequent observations have
been made in 2000 December and 2001 April during the AO2 cycle with
the same configuration as the second observation.

Over the past two years, the data reduction techniques
developed at Penn State have expanded to include new
methods for correcting the spatial and spectral degradation of the
ACIS data caused by the radiation damage, known as Charge Transfer
Inefficiency (CTI) \citep{townsley00,townsley01a}, and the use of 
charge spreading to obtain angular resolution at the sub-pixel 
level \citep{Tsunemi01}. The expected effects of the CTI correction 
include an increase of the number of detected events and improved 
event energies and the energy resolution \citep{townsley00,townsley01a}. 
By applying the ``sub-pixel resolution''
method, we expect the angular resolution to improve by $\sim$10\%
\citep{mori01}. Since the preliminary results from the first two GTO
observations as reported in B00 did not incorporate the new data
processing, we have reprocessed them for consistency with the
third and fourth observations.

We screened all four data sets with the flight timeline filter and
turned off the pixel randomization for the highest possible angular resolution.
The CTI correction was then applied before further data screenings by
status, grade, and the energy selections. The ``flaring'' pixels were
removed and the {\it ASCA} grades (02346) were selected. Photons between
0.3 keV and 8.0 keV were extracted for data analysis. The lightcurves
around the source regions were examined for possible contamination from
variable background emission and no severe variability was found.
The typical pileup fraction was small ($\la$4\%) and can be ignored.
After applying these reduction steps, the
effective exposures are 116 ks, 9 ks, 99 ks,
and 18 ks for the four observations.
We then applied the sub-pixel resolution method to improve the
angular resolution of the images.
The source position of RA = 05$^{h}$
35$^{m}$ 27$^{s}$.97, Dec = $-$69$^{\circ}$ 16$'$ 11$\farcs$09,
based on {\it Hipparcos} and VLBI data \citep{reynolds95}, was used to
register the images,
and a circular region with a 2$\arcsec$ radius was used for the extraction of
the source spectrum. The total source counts are 690, 607, 9031, and
1800 counts for observations 1, 2, 3, and 4, respectively.

The observed angular size of the SN 1987A remnant in the
optical and radio bands is only about 1$\farcs$5 and the ACIS detector
pixel size (0$\farcs$492) is not adequate to resolve the remnant.
Fortunately, the intentional ``dithering'' of the detector array on the
sky moves the supernova image across the detector pixels, allowing
us to improve the effective resolution by deconvolution techniques.
We thus apply a maximum likelihood algorithm \citep{richardson72,lucy74} 
as described in B00, using 0$\farcs$125 sky pixels
(except for the third image, where the improved photon statistics allows
us to use 0$\farcs$0625 pixels for the deconvolution).
Since the dynamic range in the detected number of photons of
our data ($\sim$600 $-$ $\sim$9000 among the four observations) is
large, we made a simple test of the reliability of the deconvolved
images by deconvolving subsets of the third observation with
different numbers of photons in the image
deconvolution process. The deconvolved subset images are consistent with
the deconvolved full image for subsets including
$\sim$400 or more counts.  We thus
conclude that all of our four images are fairly reliable
down to an angular resolution of $\sim$0$\farcs$1.
The difference in the appearance of the bright spots in the first image
compared to those in the other three can be attributed to spectral
differences between the spots and to the fact that the instrument response
for the first image is quite different than the response for the last
three due to the insertion of the HETG in the beam for the first image only.

\begin{figure}[]
\figurenum{1}
\centerline{\includegraphics[angle=-90,width=12cm]{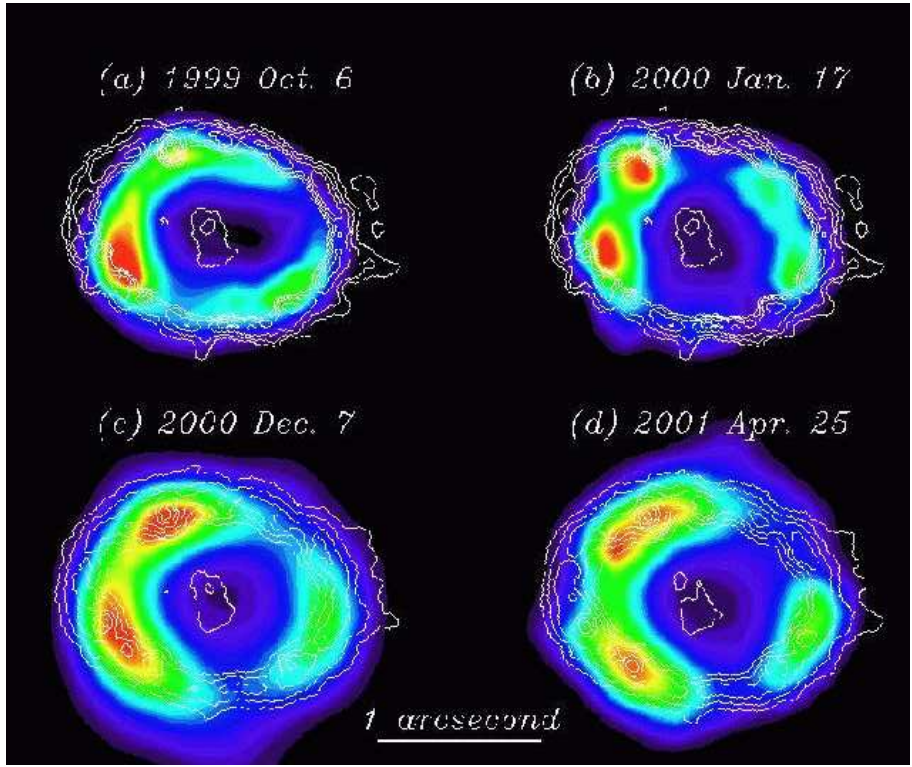}}
\figcaption[]{ACIS images for SN 1987A. (a) 1999 October 6,
(b) 2000 January 17, (c) 2000 December 7, and (d) 2001 April 25. 
The overlaid contours are from {\it HST} H${\alpha}$ images as 
taken 2000 February (a,b), 2000 November (c), and 2001 April (d).
\label{fig:fig1}}
\end{figure}

The absolute astrometry of the raw data has improved since the first
two observations. However, comparisons of the four X-ray images with the
{\it HST} images indicate that there appear to be slight aspect
errors among the {\it Chandra} observations at the level of $\sim$0$\farcs$1.
Because of the small angular size of the SN 1987A remnant ($\ga$1$\arcsec$),
this level of uncertainty in the astrometry is still significant and
we had to adjust it by hand to obtain the best
agreement between the overall X-ray and optical images.
The final images are smoothed with
$\sim$0$\farcs$1 FWHM for the presentation in Figure \ref{fig:fig1}.

\section{\label{sec:analysis} ANALYSIS \& RESULTS}

\subsection{\label{subsec:image} X-ray Images}

The four X-ray images are presented in Figure \ref{fig:fig1}, overlaid
with contours of the {\it HST} H$\alpha$ images.
Two new {\it HST} images (taken in 2000 November and
2001 April) have been kindly provided by Peter Challis and the SINS
collaboration and these new images are overlaid on the third and the fourth
{\it Chandra} images, respectively. For the first and the second
{\it Chandra} images, we use the {\it HST} image taken on 2000 February.

As reported in B00, the X-ray remnant has a shell-like morphology
with an overall brightening in the eastern half
of $\sim$20\% $-$ 30\% compared with the western half.
The brightest X-ray ``knots'' in the eastern half are generally
correlated with the optical spots as observed with {\it HST}.

\begin{figure}[]
\figurenum{2}
\centerline{\includegraphics[angle=-90,width=12cm]{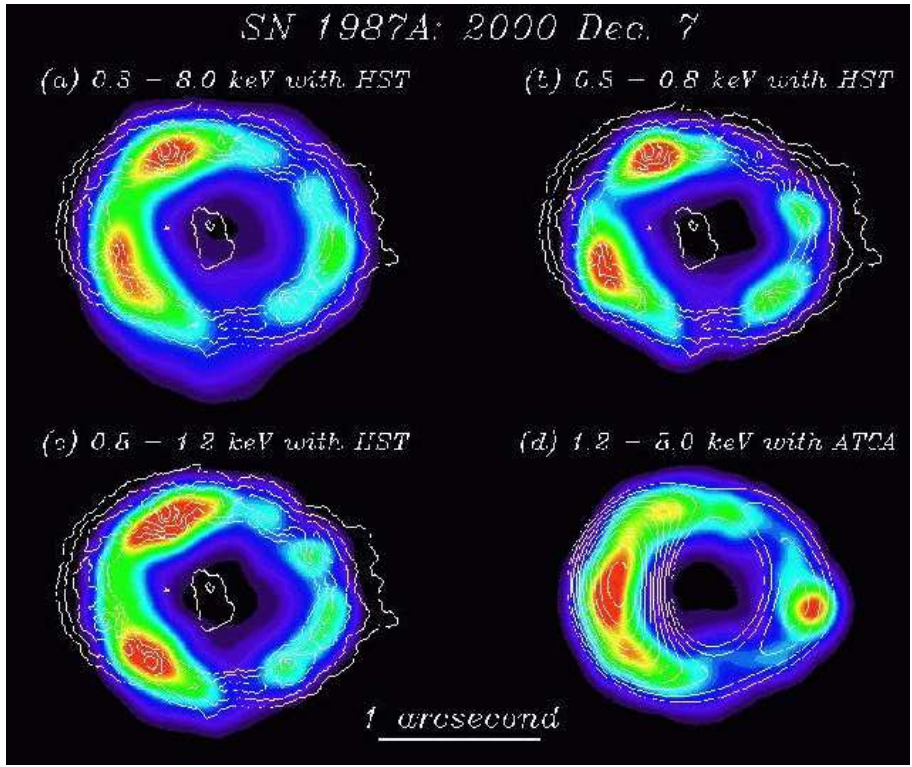}}
\figcaption[]{ACIS images of SN 1987A as taken on 2000 December 7.
(a) 0.3 $-$ 8.0 keV band, (b) 0.3 $-$ 0.8 keV, and (c) 0.8 $-$ 1.2 keV
images overlaid with the {\it HST} contours taken on 2000 November.
(d) 1.2 $-$ 8.0 keV image is overlaid with the {\it ATCA} 8 GHz contours.
\label{fig:fig2}}
\end{figure}

In order to investigate the energy-dependence of the X-ray morphology,
we have divided the X-ray images into three broad subbands: the soft
band (0.3 $-$ 0.8 keV), the mid band (0.8 $-$ 1.2 keV), and the hard
band (1.2 $-$ 8.0 keV). These three subbands are intended to include
the Oxygen line features at 0.6 $-$ 0.7 keV (soft band), Ne line at
0.9 $-$ 1 keV (mid band), and Mg and Si lines in the 1.3 $-$ 2 keV
(hard band) (see \S\ref{subsec:spec})
as well as containing relatively equal number of photons ($\sim$one-third
of the total counts in each subband). Due to the limited number of 
photons, reliable subband images are only available for the third and
the fourth observations. In Figure \ref{fig:fig2} the subband images for
the third observation are displayed; each subband image contains
$\sim$3000 counts. The pixel size in these images is 0$\farcs$0625.

Figure \ref{fig:fig2} shows that the X-ray peaks in the soft and mid band
images are well correlated with the optical spots.
Optical Spot 1 at the north-east of the optical inner
ring (position angle 29$^{\circ}$; Michael et al. 2000) is seen
as a bright X-ray knot.
Optical Spots 2 $-$ 4 at the south-east (position angles 91$^{\circ}$
$-$ 106$^{\circ}$; Lawrence et al. 2000) are also X-ray brightened.
In the western half, although the overall X-ray brightness is lower than
the east, the relatively bright X-ray emission in the south-west may
also be associated with an optical spot (spot 6 at position angle
230$^{\circ}$; Lawrence et al. 2000).
On the other hand, the hard band image, which is {\it not} well 
correlated with the optical spots, agrees very well with 
the structure seen in the radio images as observed with the
Australian Telescope Compact Array (ATCA) at 8 GHz. The hard 
band X-ray image peaks at around the mid-points of the eastern and
the western shells (position angles at $\sim$90$^{\circ}$ and
$\sim$270$^{\circ}$) (Figure \ref{fig:fig2}d). 

\begin{figure}[]
\figurenum{3}
\epsscale{1.0}
\plottwo{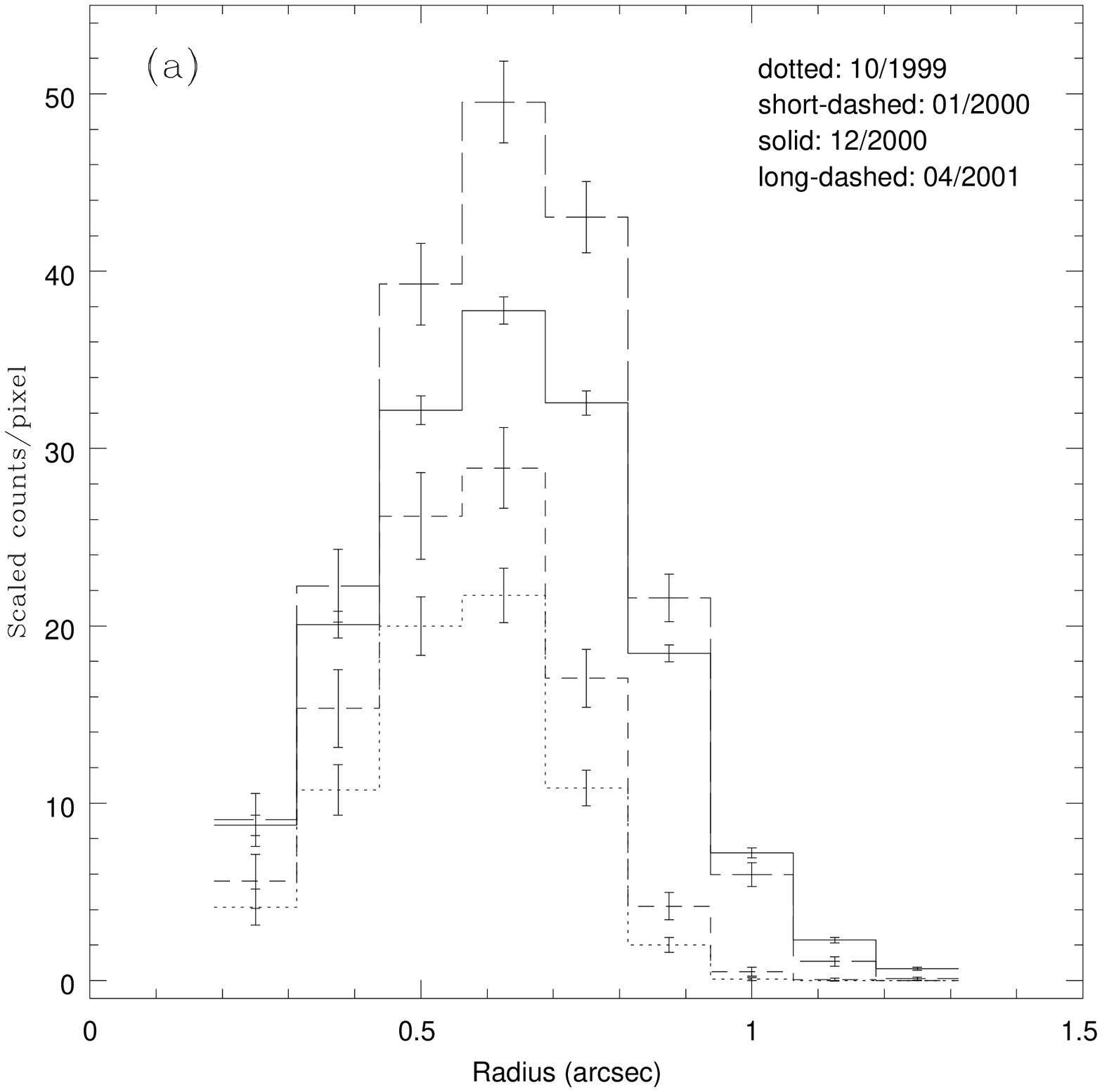}{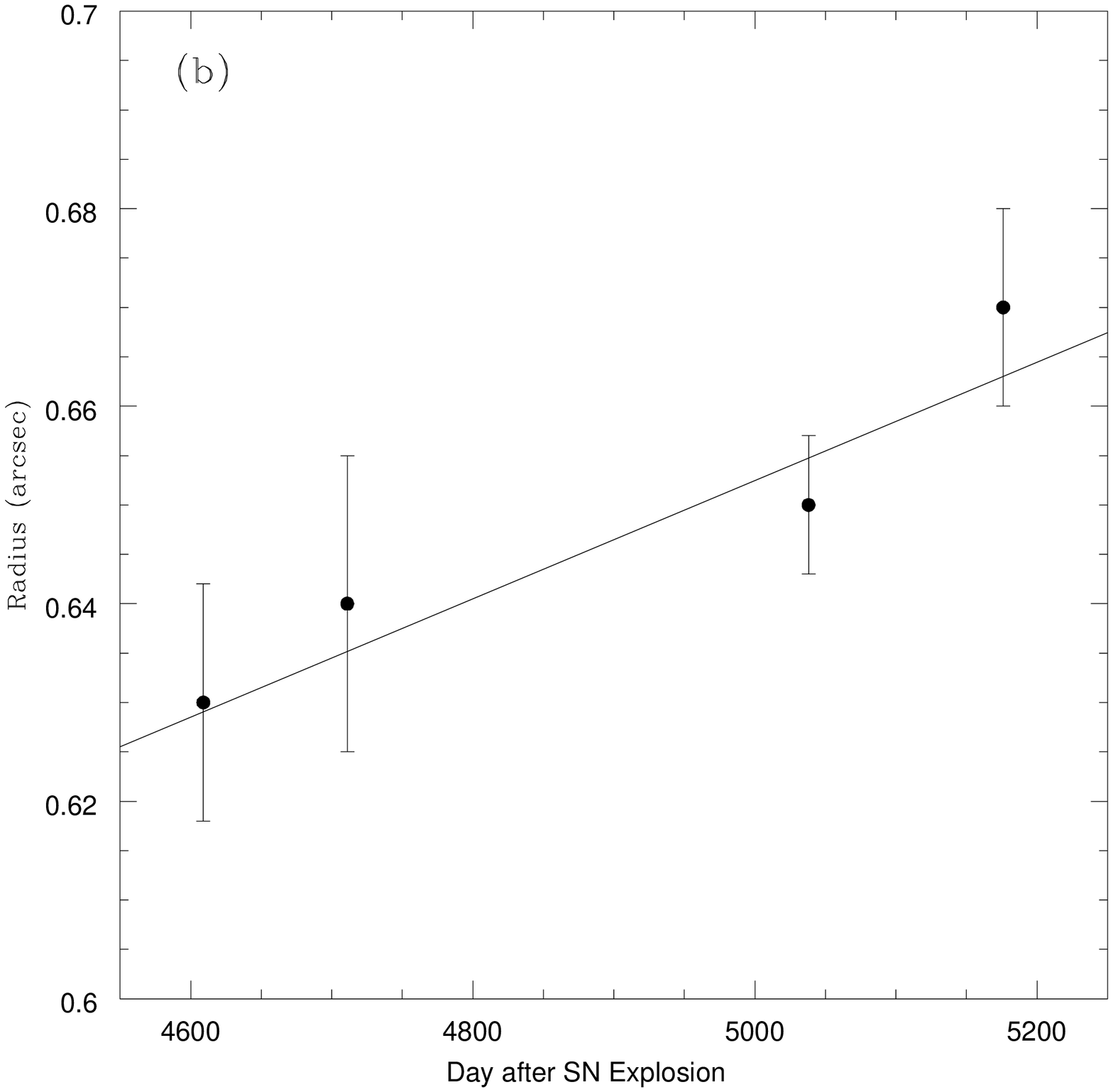}
\figcaption[]{Radial profiles of four observations of SN 1987A.
(a) is taken from 0$\farcs$125 annular regions over an $\sim$1$\arcsec$
radius around the peak X-ray intensity. (b) is the long-term variation of
the mean radius of the X-ray count distribution as obtained with a Gaussian
fit. The solid line is the best-fit linear increase rate representing
an expansion velocity of $\sim$5200 km s$^{-1}$.
\label{fig:fig3}}
\end{figure}

Considering the rapid propagation of the supernova blast wave and
the $\sim$18 month separation between the first and the fourth
observations, we investigate the possibility of detecting the radial
expansion of the X-ray remnant. Given the blast wave velocity ($\sim 
4000$ km s$^{-1}$) inferred by the hydrodynamic models \citep{borkowski97b}, 
we expect the diameter of the remnant to change by less than 
$0 \farcs 1$, and such changes will not be apparent to even
a detailed direct examination of the images.
However, we can increase our sensitivity to such effects by 
averaging over the entire remnant using radial profiles.
In this way we may lose some detailed information such as
the inclination of the ring to the line of sight and 
individual discrete features, if any, but we can make a 
statistically more reliable comparison for the overall averaged
remnant size. 
In Figure \ref{fig:fig3}a, radial profiles for the four observations 
are displayed. Each bin represents the X-ray intensity from a 
0$\farcs$125 width annular region centered on the adopted source position.
The peak bin is at a radius of $\sim$0$\farcs$63 for all four 
observations. The brightness at the radius of $\sim$0$\farcs$75 
however appears continuously increasing while the intensity at 
$\sim$0$\farcs$50 bin decreases such that the mean radius of the 
shell moves measurably outwards with time.
In order to accurately quantify this effect, we reconstructed the
radial profiles centered on the mean position of the count
distribution (or ``center of mass'') with a smaller bin (0$\farcs$05
annular region). We then fit these radial profiles with Gaussians 
and plot the best-fit Gaussian peaks versus time in Figure~\ref{fig:fig3}b. 
The peak radius increases by $\sim 0 \farcs 04$ from 1999 October 
to 2001 April. We tested this estimation with different annular
bin sizes of 0$\farcs$025, 0$\farcs$05, and 0$\farcs$125 because 
such a small change in angular size of 0$\farcs$04 might have been
affected by the selected bin size. The results are the same regardless
of the selected annular bin sizes. We also compared the averaged
angular sizes by simply estimating the mean (or ``center of mass'')
radii among the observations instead of fitting with a Gaussian
in case of the existence of any systematic bias with the Gaussian
fittings. Regardless of the annular bin sizes, the $\sim$0$\farcs$04 
increase in the radius is persistent with this simple estimation as well. 

The best-fit expansion rate of the X-ray remnant, indicated by 
the solid line in Figure~\ref{fig:fig3}b, is 5200 $\pm$ 2100 
km s$^{-1}$. This expansion rate is in agreement with the rate 
determined from radio observations \citep{gaensler00}, and is also 
consistent with the {\it HST} STIS observations of Ly$\alpha$ 
and H$\alpha$ \citep{michael98} as well as with the theoretical
predictions \citep{borkowski97b}. Finally, we measured the radial 
expansions along the major (east-west) and minor (north-south) 
axes of the ring in order to probe the effects of the inclination 
angle of the ring to the line of sight in the overall expansion 
estimations.
The results appear generally consistent with the overall expansion: 
i.e., the radial expansion is $\sim$0$\farcs$03 along both 
directions of the major and minor axes. 
The slight difference in the expansion rate between the major and 
minor axis directions is statistically insignificant.

\subsection{\label{subsec:spec} Spectrum \& Lightcurve}

\begin{figure}[]
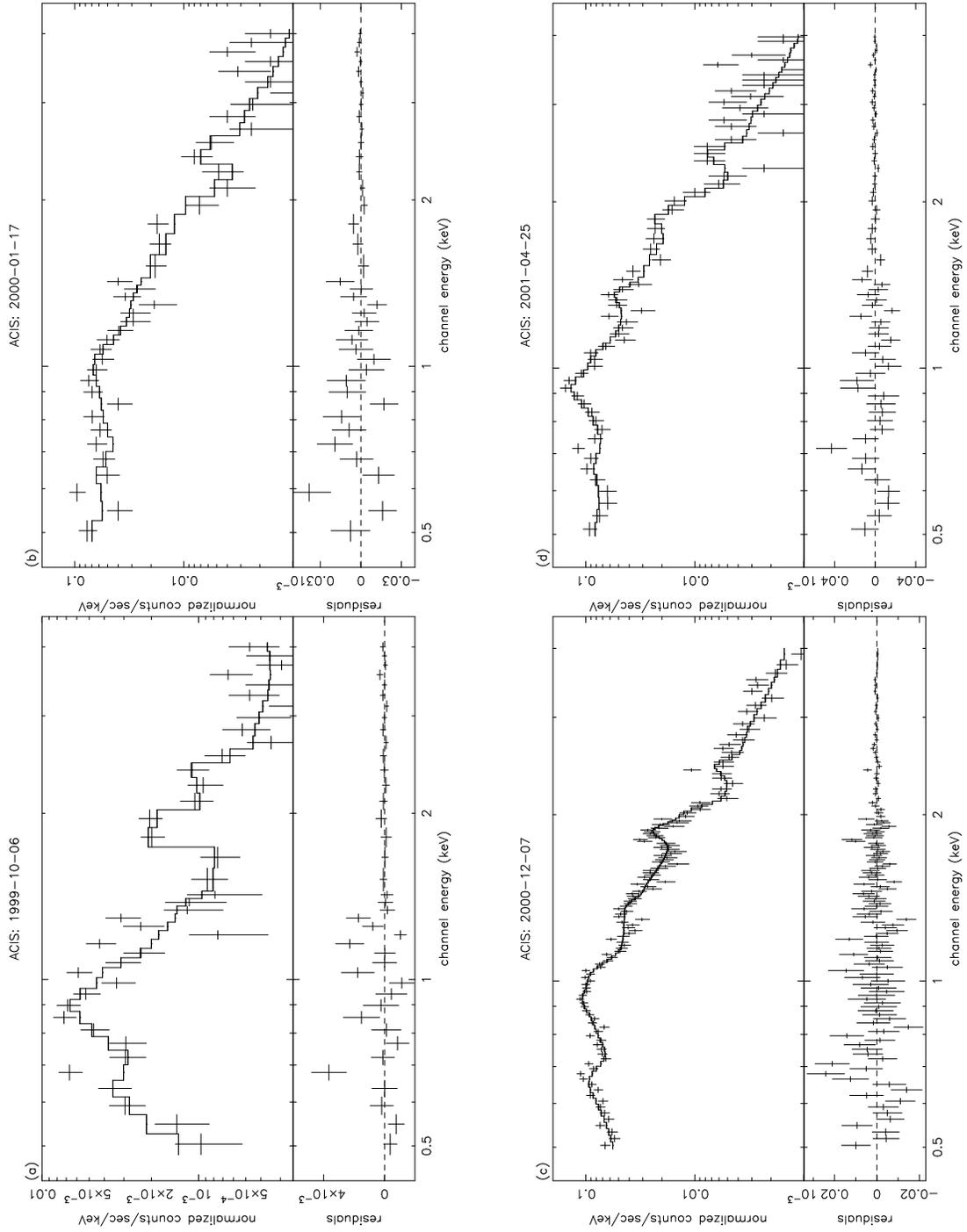

\figurenum{4}
\epsscale{0.9}
\plottwo{fig4b.ps}{fig4d.ps}

\epsscale{2.0}
\plottwo{fig4a.ps}{fig4c.ps}
\figcaption[]{Spectrum of SN 1987A in the 0.5 $-$ 4.0 keV band.
\label{fig:fig4}}
\end{figure}

The undispersed spectra of SN 1987A from all four {\it Chandra} 
observations are presented in Figure \ref{fig:fig4}.
Each spectrum has been rebinned to provide a minimum of 20
counts per bin. The energy range of 0.5 $-$ 4.0 keV
(which typically contains $\ga$90\% of the total counts) is
used for the spectral fitting. For the spectral analysis of
our CTI corrected data, we have utilized the response matrices 
appropriate for the spectral redistribution of the CCD,
as generated at Penn State \citep{townsley01b}. 
Broad emission line features are evidently present in each 
observation, supporting a thermal origin of the observed X-ray 
emission. The broad line profiles correspond to O, Ne, Mg, and 
Si line complexes as reported in B00 from the dispersed spectrum 
of the first observation.

The observed spectrum can be described with a plane-parallel shock
model with an electron temperature of $kT$ = 2 $-$ 4 keV and an average
ionization timescale of $nt$ $\sim$ 8 $\times$ 10$^{10}$ cm$^{-3}$ s
being absorbed by N$_{H}$ $\sim$ 1 $\times$ 10$^{21}$ cm$^{-2}$.
The elemental abundances were fixed for H (= 1), He (= 2.57), and C 
(= 0.09) at the appropriate values for the inner circumstellar ring 
\citep{lundqvist96} and Ca (= 0.34) and Ni (= 0.62) at values for 
the LMC \citep{russell92} (hereafter, all abundances are with respect
to the solar) since the contribution from these species in 
the spectral fitting is expected to be insignificant in the selected 
energy range. Other elements were allowed to vary and fit the data 
with sub-solar abundances: e.g., for the 2000-December data, the 
best-fit abundances are N = 0.01, O = 0.04, Ne = 0.12, Mg = 0.08, 
Si = 0.24, S = 0.40, and Fe = 0.07. Our simple model consistently 
fits all four data sets with moderately acceptible statistics 
($\chi^{2}_{\nu}$ = 1.2 $-$ 1.6).

\begin{figure}[]
\figurenum{5}
\centerline{\includegraphics[angle=90,width=12cm]{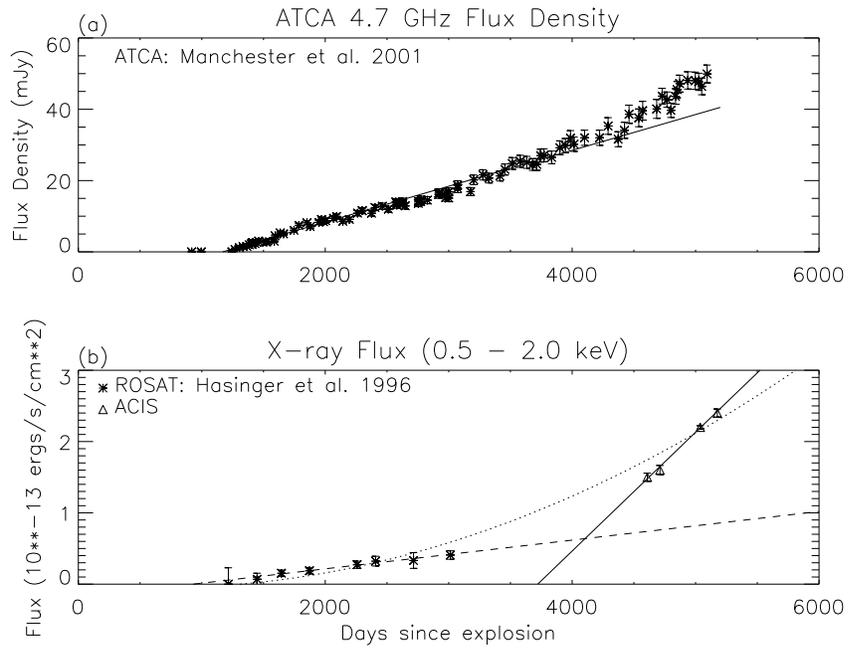}}
\figcaption[]{{\it Top panel}: (a) The long-term lightcurve of SN 1987A
in the 4.7 GHz (ATCA). {\it Bottom panel}: (b) The long-term X-ray lightcurve
of SN 1987A ({\it ROSAT} and {\it Chandra} ACIS).
\label{fig:fig5}}
\end{figure}

With this simple model, the 0.5 $-$ 2.0 keV band X-ray fluxes of
SN 1987A for our four observations are $\sim$1.5
$\times$ 10$^{-13}$ ergs s$^{-1}$ cm$^{-2}$, $\sim$1.6 $\times$
10$^{-13}$ ergs s$^{-1}$ cm$^{-2}$, $\sim$2.2 $\times$ 10$^{-13}$
ergs s$^{-1}$ cm$^{-2}$, and $\sim$2.4 $\times$ 10$^{-13}$ egrs
s$^{-1}$ cm$^{-2}$.
(The X-ray fluxes for the first and the second observations have been
recalculated using our current processing with CTI correction,
and the revised fluxes increased by $\sim$15\% from those
presented in B00.) Based on these flux estimations,
the long-term X-ray lightcurve of SN 1987A is presented in Figure
\ref{fig:fig5}. In Figure \ref{fig:fig5}a, an updated long-term
radio flux variation \citep{manchester01} is presented for 
comparison. For the X-ray lightcuve, the {\it ROSAT} fluxes were
taken from Hasinger et al. (1996) and have been converted for 
comparison with the ACIS data as presented in B00 (Figure
\ref{fig:fig5}b). The 0.5 $-$ 2.0 keV X-ray flux and luminosity
between 1999 and 2001 are listed in Table \ref{tbl:tab2}.

The X-ray flux has been increasing at a constant rate for the past 
18 months, and is now brightening much faster than expected
from the linear extrapolation of the {\it ROSAT} lightcurve.
We note that the interpolated date of this slope change in the 
lightcurve, for both the X-ray and radio data, is about 1997, when 
the first optical spot was discovered. The net
increase in the X-ray flux between 1999 October and 2001 April
is $\sim$60\%. Assuming a distance of 50 kpc, and correcting
for the interstellar absorbing column of $10^{21}$ cm$^{-2}$ 
inferred from our spectral fits, the 0.5 $-$ 2 keV X-ray
luminosity was $\sim$0.7 $\times$ 10$^{35}$ ergs s$^{-1}$ ($\sim$1.0
$\times$ 10$^{35}$ ergs s$^{-1}$, in the 0.5 $-$ 10 keV band)
in 1999 October and was $\sim$1.0 $\times$ 10$^{35}$ ergs s$^{-1}$
($\sim$1.3  $\times$ 10$^{35}$ ergs s$^{-1}$, in the 0.5 $-$ 10 keV band)
in 2001 April. 

Assuming a spherical shell for the X-ray emitting 
volume, the best-fit emission measure for the 2000-December 
observation implies a postshock electron density of $n_e$ $\sim$210 
cm$^{-3}$ to $\sim$420 cm$^{-3}$.
In this estimation, we have assumed a spherical shell of
an inner radius of 0$\farcs$6 \citep{gaensler00} and a range of the 
outer radius of 0$\farcs$7 $-$ 0$\farcs$9 (from our data) for 
the X-ray emitting volume. We also assumed $n_e$ $\sim$ 1.5$n_H$
for the ring abundances (e.g., Masai \& Nomoto 1994). 
Considering our simple modeling and the embedded uncertainty 
in the assumed geometry, the derived range of the electron density 
is in good agreement with the previously suggested values for the
preshock \hii region \citep{chevalier95,borkowski97b,lundqvist99}, 
assuming density enhancement by a factor of 4 at the front of a strong 
adiabatic shock.


\section{\label{sec:disc} DISCUSSION}

The overall appearance of the X-ray images of SN 1987A shows little
change over the one and a half year observation period: i.e., the
shell-like overall morphology, brighter emission in the eastern half,
and the general correlations of the X-ray brightening with the optical
spots.
(The most noticeable difference is the supression of Spot 1 in our first
observation, which we believe to be due to lower sensitivity to this soft
feature, due to the insertion of the HETG into the optical path for this
observation.)
The correlations of the X-ray peaks with the optical spots in the 
images of the first and the second observations are generally better
than found by B00, which is likely an effect from our improved data 
processing.
The observed shell-like X-ray emission is interpreted as X-ray
emission from the shocked SN ejecta and the shocked circumstellar
material between the supernova blast wave and the reverse shock
(B00 and the references therein). The origin of the overall X-ray 
brightness asymmetry between the east and the west may be related
to the asymmetric distribution of SN ejecta and/or to the density 
variation of the circumstellar medium. Extensive spectral analyses
will be necessary in order to answer this question, which is beyond
the scope of the current work. 

The comparisons among the images and the radial profiles show 
evidence of the radial expansion of the X-ray shell of SN 1987A.
We have assumed constant expansion during the 18 month duration of
our monitoring observations. 
Although the blast wave velocity may not be constant in time
due to the interaction with the circumstellar material, 
the quality of the data do not justify more complex fits at this time.
The change in the angular radius ($\sim$0$\farcs$04) is  
smaller than the angular resolution of the deconvolved images.
Although the peak radius appears to be monotonically increasing,
the derived expansion rate is only a $2.5 \sigma$ result.
Considering systematic uncertainties associated with
the image processing (e.g., the image deconvolution, the selected 
source position etc.), we therefore must consider
this result to be of marginal significance.
Follow up observations will be necessary for more definitive
measurements of this interesting aspect of the SNR.

The broad subband images show that the X-ray bright knots in the soft
band are well correlated with the optical spots, typically
within $\sim$0$\farcs$05, while such correlations are not observed in
the hard band. Particularly the origin of the bright X-ray emission
feature at position angles $\sim$90$^{\circ}$ and $\sim$270$^{\circ}$
in the hard band draws our attention. We have investigated the possibility
of a coincidental detection of a background extragalactic
object. We searched for possible counterparts in 33
multi-wavelength catalogs of galaxies and radio sources available
through the HEASARC on-line database and no such candidates were found
within 1$'$ radius of our reference source position. We have estimated
the ``source'' flux from the spectrum of the bright hard X-ray knot
in the west ring (position angle $\sim$ 270$^{\circ}$). The extracted
spectrum contains $\sim$400 photons and is best fitted in the $0.5 - 5$ 
keV band with a power law plus plane-parallel shock model ($\Gamma$ = 
2.1, $kT$ = 2 keV, $nt$ = 9 $\times$ 10$^{10}$ cm$^{-3}$ s,
$\chi^{2}_{\nu}$ $\sim$ 1.0). The estimated flux is $\sim$1 $\times$
10$^{-14}$ ergs s$^{-1}$ cm$^{-2}$ in both of the 0.5 $-$ 2 keV and
the 2 $-$ 10 keV band, and based on the recent {\it Chandra} logN-logS 
relations of the {\it Hubble Deep Field} \citep{brandt01},
the probability of a coincidental detection of an extragalactic
source within the angular size of SN 1987A ($\sim$1 arcsec$^{2}$)
is only $\sim$10$^{-5}$. The spectral index ($\Gamma$ = 2.1) may be
in the range of nearby broad line AGNs but clearly deviates from
typical AGNs ($\Gamma$ $\la$ 1.5) (Alexander et al. 2001 and references
therein). We thus conclude that this hard X-ray spot is emission 
from the supernova remnant.
These bright hard X-ray knots are on the other hand well correlated
with the radio emission.
This can be understood if the hard X-ray emission and radio emission
both originate in the fast shock ($\sim$ 4000 km s$^{-1}$)
propagating into the \hii region.
As this shock hits the denser, neutral ring material, the shock slows
down significantly.  Soft X-rays are emitted from the leading edge of
these knots, with oblique radiative shocks on the sides of the dense
knots ($\la$ 300 km s$^{-1}$) providing the correlated optical and UV 
emission \citep{michael00a, michael00b, pun01}.

A recent result from the observation of SN 1987A with {\it XMM-Newton
Observatory} has suggested a non-thermal component at $>$ 4 keV,
which is speculated to be emission from the embedded pulsar and
its wind nebula \citep{aschenbach01}.
Although our spectral analysis did not require a power-law component,
the observed correlation between the hard X-ray and radio images
may suggest such a contribution. We used the same one
temperature plane-parallel shock model, now with the insertion of a
power law, in the 0.5 $-$ 6 keV band to investigate this possibility.
The improvement of the fit
after including a power-law component appears statistically
significant based on the F-test. The overall fits are however
practically indistinguishable, e.g., ${\chi}^{2}_{\nu}$ = 1.05
with a power law and ${\chi}^{2}_{\nu}$ = 1.12 without a power law,
for the purposes of our preliminary spectral analysis.
There are small changes in the best-fit electron temperature and the
ionization timescale, which are insignificant. The photon index for
the best-fit power law ($\Gamma$ $\sim$ 2.3) is consistent with the
shock-accelerated synchrotron radiation observed with young Galactic
supernova remnants \citep{koyama95,koyama97,keohane97,allen97}.
The inclusion of a power law results in relatively higher elemental
abundances, generally at above-solar level compared with the sub-solar
abundances without the power-law component. 
This simple test implies that the presence of a power-law 
component as a contributor in the observed X-ray spectrum, 
particularly in the hard band, cannot be ruled out, although it is 
not required by our data. The physical origin of the non-thermal 
emission, if detected at all, is however unclear considering the 
early phase of the SNR evolution: i.e., it is unlikely from the 
embedded pulsar or its nebula since we have yet to see such evidence 
with our high resolution ACIS images (see discussion below), and 
the age of the SNR may not be old enough to accelerate the particles 
to hard X-ray energies, although the best-fit photon index is 
plausible for such origins. The implied high metal abundances 
with the power-law component suggest the SN ejecta as a dominating 
source of the X-ray emission, which may also be inappropriate for
such an early stage of the SNR as predicted by hydrodynamic models
\citep{borkowski97b}. We thus defer
the issue of the non-thermal contribution in the observed spectrum of 
SN 1987A to future investigations.

The best-fit parameters for the observed spectrum indicate no significant
change from the results in B00. The best-fit electron temperatures
are substantially lower than the expected postshock ion temperature
for the implied shock velocities, which is not surprising since the 
time scale to reach an equilibrium between the electron and ion 
temperatures is much longer than the age of SN 1987A. This 
non-equilibrium status between the electron and ion temperatures
suggests that it may require more complicated treatments in the 
spectral analysis instead of a simple one temperature plane-parallel 
shock model as we have utilized in the current work.
Due to the limited number of photons, an extensive spectral analysis
is feasible only with the third observation and such an analysis
is presented in a separate paper \citep{michael01}. Michael et al. (2001)
demonstrate that the spectrum of SN 1987A can be best described with
more complex models rather than the simple model used here.

The X-ray lightcurve demonstrates that the remnant of SN 1987A has
been continuously brightening in X-rays over the last $\sim$10 years.
Between 1991 and 1995 the 0.5 $-$ 2.0 keV band X-ray flux was
increasing at a rate of $\sim$2.0 $\times$ 10$^{-17}$ ergs
s$^{-1}$ cm$^{-2}$ per day (${\chi}^{2}_{\nu}$ = 0.1). 
As of 2001 April, the observed intensity
of the X-ray remnant has increased by a factor of $\sim$6 since the
1995-November {\it ROSAT} observation. This is almost 3 times brighter than
would be expected from the extrapolations of the flux increase rate
with the {\it ROSAT} observations. The 0.5 $-$ 2.0 keV X-ray flux
is currently increasing at a constant rate of $\sim$1.67 $\times$
10$^{-16}$ ergs s$^{-1}$ cm$^{-2}$ per day (${\chi}^{2}_{\nu}$ = 0.5), 
which is $\sim$8 times higher rate than it was in 1995.
The combined {\it ROSAT} and {\it Chandra} data cannot be fitted by
a quadratic light curve (${\chi}^{2}_{\nu}$ = 6.3), although 
suggested on theoretical grounds \citep{mn94,hasinger96} 
(see dotted line in Figure~\ref{fig:fig5}b).
A simple power law with an index of 2.8 provides a better fit
to the combined lightcurve, but cannot fit the {\it Chandra} data 
points (${\chi}^{2}_{\nu}$ = 3.5).

The observed bright X-ray emission near the optical spots and
the continuously increasing X-ray flux at a higher rate most likely imply
that the supernova blast wave is closely approaching the inner ring
and that we may shortly be able to see the dramatic turn-up in the
remnant luminosity when the blast wave finally reaches and sweeps through
the dense inner ring.

As of 2001 April, we still observe no direct evidence of a point source
within the X-ray remnant of SN 1987A. We thus estimated a point source
detection limit by performing a Monte Carlo simulation to add
a simulated point source at the center of the observed SN 1987A remnant.
The simulated point source was generated
by convolving point sources of various fluxes with the {\it Chandra} point
spread function with photon statistics. The simulated images (with the
point source) were compared with the observed image using a $\chi$$^{2}$
test. We have utilized the 2000-December observation for this purpose
since the large number of detected photons in this observation would
provide the most reliable estimation.
Because the ejecta are still expected to be optically thick at soft X-ray
energies, we restricted this test to hard X-rays ($E > 2$ keV).
The 90\% confidence limit on the
point source counts is $\sim$13\% of the total counts at $>$ 2 keV,
which implies an {\it observed} upper limit of
$\sim$5.5 $\times$ 10$^{33}$ ergs
s$^{-1}$ in the 2 $-$ 10 keV luminosity for any embedded point source.
Since we do not know the optical depth of the overlaying ejecta, we cannot
place limits directly on the actual X-ray emission from the
putative compact object.

\section{\label{sec:summary} SUMMARY AND CONCLUSIONS}

Using the high angular resolution of the {\it Chandra X-ray Observatory},
we have resolved the X-ray remnant of SN 1987A in four different observations
taken over an 18 month period.	The stability of the features lends
confidence in our image processing techniques, which produce an effective
resolution of $\sim$0$\farcs$1. The X-ray flux is increasing $\sim 8$
times faster now than it was five years ago.
Soft X-ray images show good agreement between
X-ray and optical bright spots, while hard X-ray images show
better correlation with radio images from ATCA.  This can be understood
in terms of a model in which a fast shock propagating into the circumstellar
\hii region produces the hard X-ray and radio emission, while slower shocks
propagating into the dense inner ring are responsible for the soft
X-ray and optical spots. Although we show that the observed X-ray 
spectrum can be described with a single temperature plane-parallel 
shock model, a more complex model, that is in general consistent with 
our simple model presented in this work, provides a more physically
plausible fit, as shown by Michael et al. (2001). Finally, we demonstrate 
that the X-ray images suggest an expansion velocity of $5200 \pm 2100$ 
km s$^{-1}$ with marginal significance, in agreement with the 
radio and optical observations as well as the theoretical models.

\acknowledgments

The authors thank L. Townsley and colleagues in the department of
Astronomy \& Astrophysics at Penn State University for developing
the software for the CTI correction and for generating correspondent
response files, which we utilized in the spectral analysis of our data. 
We also thank K. Mori for providing the codes to perform the 
subpixel-resolution and P. Challis and the SINS collaboration for 
providing the {\it HST} images.
S.P. thanks K. Lewis, K. Mori, F. Bauer, D. Alexander, and C. Vignali
for their kind help and valuable discussion.
This work was funded by NASA under contract NAS8-3852 and by SAO
under grant GO1-2064B.

\clearpage

\begin{deluxetable}{cllcc}
\footnotesize
\tablecaption{List of the {\it Chandra} Observations of SN 1987A
\label{tbl:tab1}}
\tablewidth{0pt}
\tablehead{\colhead{Observation ID} & \colhead{Date (Age)\tablenotemark{a}}
& \colhead{Instrument} & \colhead{Exposure (ks)} &
\colhead{Source Counts\tablenotemark{b}}}
\startdata
00124+01387\tablenotemark{c} & 1999 October 6 (4609) & ACIS-S3 + HETG & 116 & 690 \\
00122 & 2000 January 17 (4711) & ACIS-S3 & 9 & 607 \\
01967 & 2000 December 7 (5038) & ACIS-S3 & 99 & 9031 \\
01044 & 2001 April 25 (5176) & ACIS-S3 & 18 & 1800 \\
\enddata

\tablenotetext{a}{Day after the SN explosion in the parentheses.}
\tablenotetext{b}{The source counts were accumulated from a circular
region with a radius of 2$\arcsec$ at 0.3 $-$ 8.0 keV.}
\tablenotetext{c}{The first observation on 1999 October was split
into two sequential observations, which were combined in the analysis.}

\end{deluxetable}

\begin{deluxetable}{ccc}
\footnotesize
\tablecaption{The 0.5 $-$ 2.0 keV Flux and Luminosity of SN 1987A
from {\it Chandra} ACIS
\label{tbl:tab2}}
\tablewidth{0pt}
\tablehead{\colhead{Day\tablenotemark{a}} & \colhead{Observed Flux} &
\colhead{Luminosity\tablenotemark{b}} \\
 & \colhead{(10$^{-13}$ ergs s$^{-1}$ cm$^{-2}$)} &
\colhead{(10$^{34}$ ergs s$^{-1}$)}}
\startdata
4609 & 1.53$\pm$0.06 & 6.5 \\
4711 & 1.57$\pm$0.06 & 6.5 \\
5038 & 2.23$\pm$0.02 & 9.1 \\
5176 & 2.44$\pm$0.06 & 10.0 \\
\enddata

\tablenotetext{a}{Day after the SN explosion.}
\tablenotetext{b}{Luminosities after removing the absorption.}

\end{deluxetable}

\clearpage

\clearpage



\clearpage

\clearpage

\begin{thebibliography}{}

\bibitem[Allen et al. 1997]{allen97} Allen, G. E., Keohane, J. W.,
Gitthelf, E. V., Petre, R., \& Jahoda, K. 1997, ApJ, 487, L97

\bibitem[Alexander et al. 2001]{alexander01} Alexander, D. M.,
Brandt, W. N., Hornschemeier, A. E., Garmire, G. P., Schneider, D. P., \&
Bauer, F. E. 2001, AJ, in press

\bibitem[Andreani et al. 1987]{andreani87} Andreani, P., Ferlet, R.,
\& Vidal-Madjar, A. 1987, Nature, 326, 770

\bibitem[Aschenbach et al. 2001]{aschenbach01} Aschenbach, B., Hasinger, G.
 et al. 2001, A\&A, Submitted

\bibitem[Blondin \& Lundqvist 1993]{blondin93} Blondin, J. M., \& 
Lundqvist, P. 1993, ApJ, 405, 337 

\bibitem[Borkowski et al. 1997a]{borkowski97a} Borkowski, K. J., Blondin,
J. M., \& McCray, R. 1997a, ApJ, 477, 281

\bibitem[Borkowski et al. 1997b]{borkowski97b} Borkowski, K. J., Blondin,
J. M., \& McCray, R. 1997b, ApJ, 476, L31

\bibitem[Bouchet et al. 2000]{bouchet00} Bouchet, P., Lawrence, S.,
Crotts, A., Sugerman, B., Uglesich, R., \& Heathcote, S. 2000, IAUC, 7354

\bibitem[Brandt et al. 2001]{brandt01} Brandt, W. N. et al. 2001, AJ, in press

\bibitem[Burrows et al. 1995]{burrows95} Burrows, C. J., Krist, J.,
Hester, J., Sahai, R., Trauger, J. T., Stapelfeldt, K. R., Gallagher III,
J. S., Ballester, G. E., Casertano, S., Clarke, J. T., Crisp, D.,
Evans, R. W., Griffiths, R. E., Hoessel, J. G., Holtzman, J. A.,
Mould, J. R., Scowen, P. A., Watson, A. M., \& Westphal, J. A. 1995,
ApJ, 452, 680

\bibitem[Burrows et al. 2000]{burrows00} Burrows, D. N., Michael, E.,
Hwang, U., McCray, R., Chevalier, R. A., Petre, R., Garmire, G. P.,
Holt, S. S., \& Nousek, J. A. 2000, ApJ, 543, L149: B00


\bibitem[Chevalier 1992]{chevalier92} Chevalier, R. A. 1992, Nature, 355, 691

\bibitem[Chevalier \& Dwarkadas 1995]{chevalier95} Chevalier, R. A., \&
Dwarkadas, V. V. 1995, ApJ, 452, L45


\bibitem[Gaensler et al. 2000]{gaensler00} Gaensler, B. M., Manchester,
R. N., Staveley-Smith, L., Wheaton, V., Tzioumis, A. K., Reynolds, J. E., 
\& Kesteven, M. J. 2000, Asymmetrical Planetary Nebulae II: From Origins 
to Microstructures, ASP Conference Series, Vol. 199. ed. by J. H. Kastner, 
N. Soker, \& S. Rappaport, p. 449 


\bibitem[Garnavich et al. 1997]{garnavich97} Garnavich, P., Kirshner, R.,
\& Challis, P. 1997, IAUC, 6710

\bibitem[Garnavich et al. 2000]{garnavich00} Garnavich, P., Challis, P.,
\& Kirshner, R. 2000, IAUC, 7360

\bibitem[Hasinger et al. 1996]{hasinger96} Hasinger, G., Aschenbach, B.,
\& Tru\"mper, J. 1996, A\&A, 312, L9

\bibitem[Keohane et al. 1997]{keohane97} Keohane, J. W., Peter, R.,
Gotthelf, E. V., Ozaki, M., \& Koyama, K. 1997, ApJ, 484, 350

\bibitem[Kirshner et al. 1987]{kirshner87} Kirshner, R. P., Sonneborn, G.,
Crenshaw, D. M., \& Nassiopoulos, G. E., 1987, ApJ, 320, 602

\bibitem[Koshiba et al. 1987]{koshiba87} Koshiba, M. et al. 1987, IAUC, 4338

\bibitem[Koyama et al. 1995]{koyama95} Koyama, K., Peter, R., Gotthelf,
E. V., Hwang, U., Matsuura, M., Ozaki, M., \& Holt, S. S. 1995, Nature,
378, 255

\bibitem[Koyama et al. 1997]{koyama97} Koyama, K., Kinugasa, K., Matsuzaki,
K., Nishiuchi, M., Sugizaki, M., Torii, K., Yamauchi, S., \& Aschenbach,
B. 1997, PASJ, 49L, 7

\bibitem[Lawrence et al. 2000]{lawrence00} Lawrence, S. S., Sugerman, B. E.,
Bouchet, P., Crotts, A. P. S., Uglesich, R., \& Heathcote, S. 2000, ApJ,
537, L126

\bibitem[Lucy 1974]{lucy74} Lucy, L. B. 1974, AJ, 79, 745

\bibitem[Lundqvist \& Fransson 1996]{lundqvist96} Lundqvist, P., \&
Fransson, C. 1996, ApJ, 464, 924

\bibitem[Lundqvist 1999]{lundqvist99} Lundqvist, P. 1999, ApJ, 511, 389

\bibitem[Luo \& McCray 1991]{luo91} Luo, D. \& McCray, R. 1991, ApJ, 379, 659

\bibitem[Luo et al. 1994]{luo94} Luo, D., McCray, R., \& Slavin, J.
1994, ApJ, 430, 264

\bibitem[Maran et al. 2000]{maran00} Maran, S., Pun, C. S. J., \&
Sonneborn, G. 2000, IAUC, 7359

\bibitem[Masai \& Nomoto 1994]{mn94} Masai, K., \& Nomoto, K. 1994,
ApJ, 424, 924

\bibitem[Manchester et al. 2001]{manchester01} Manchester, R. et al. 2001,
PASA, Submitted

\bibitem[Michael et al. 1998]{michael98} Michael, E., McCray, R.,
Pun, C. S. J., Borkowski, K., Garnavich, P., Challis, P., Kirshner, R. P.,
Chevalier, R., Filippenko, A. V., Fransson, C., Panagia, N., Phillips,
M., Schmidt, B., Suntzeff, N., \& Wheeler, J. C. 1998, ApJ, 509, L117

\bibitem[Michael et al. 2000]{michael00a} Michael, E., McCray, R.,
Pun, C. S. J., Garnavich, P., Challis, P., Kirshner, R. P., Raymond, J.,
Borkowski, K., Chevalier, R., Filippenko, A. V., Fransson, C., Lundqvist,
P., Panagia, N., Phillips, M. M., Sonneborn, G., Suntzeff, N. B., Wang,
L., \& Wheeler, J. C. 2000, ApJ, 542, L53

\bibitem[Michael 2000]{michael00b} Michael, E. 2000, ApJS, 127, 429

\bibitem[Michael et al. 2001]{michael01} Michael, E. et al. 2001, ApJ, 
Submitted

\bibitem[Mori et al. 2001]{mori01} Mori, K., Tsunemi, H., Miyata, E.,
Baluta, C. J., Burrows, D. N., Garmire, G. P., \& Chartas, G. 2001,
``New Century of X-ray Astronomy'', March 6 $-$ 8, 2001, Yokohama, Japan

\bibitem[Pun et al. 1997]{pun97} Pun, C. S. J., Sonneborn, G., Bowers, C.,
Gull, T., Heap, S., Kimble, R., Maran, S., \& Woodgate, B. 1997, IAUC, 6665

\bibitem[Pun et al. 2001]{pun01} Pun, C. S. J., Michael, E., Zhekov, S. A.,
McCray, R. et al. 2001, ApJ, Submitted

\bibitem[Reynolds et al. 1995]{reynolds95} Reynolds, J. E. et al. 1995,
A\&A, 304, 116

\bibitem[Richardson 1972]{richardson72} Richardson, W. H., 1972, J. Opt. Soc.
Am., 62, 55

\bibitem[Russell \& Dopita 1992]{russell92} Russell, S. C. \& Dopita,
M. A. 1992, ApJ, 384, 508

\bibitem[Shelton 1987]{shelton87} Shelton, I. 1987, IAUC, 4316

\bibitem[Sonneborn et al. 1987]{sonneborn87} Sonneborn, G., Altner, B.,
\& Kirshner, R. P. 1987, ApJ, 323, L35

\bibitem[Townsley et al. 2000]{townsley00} Townsley, L. K., Broos, P. S.,
Garmire, G. P., \& Nousek, J. A. 2000, ApJ, 534, L139

\bibitem[Townsley et al. 2001a]{townsley01a} Townsley, L. K., Broos, P. S.,
Nousek, J. A., \& Garmire, G. P. 2001a, Nuclear Instruments \& Methods in
Physics Research Section A, in press

\bibitem[Townsley et al. 2001b]{townsley01b} Townsley, L. K., Broos, P. S.,
Chartas, G., Moskalenko, E., Nousek, J. A., \& Pavlov, G. G. 2001b,
Nuclear Instruments  \& Methods in Physics Research Section A, in press


\bibitem[Tsunemi et al. 2001]{Tsunemi01} Tsunemi, H., Mori, K., Miyata, E.,
Baluta, C., Burrows, D. N., Garmire, G. P., \& Chartas, G. 2001, ApJ, 554,
496

\bibitem[Wang \& Mazzali 1992]{wang92} Wang, L. \& Mazzali, P. A. 1992, 
Nature, 335, 58

\bibitem[Weisskopf et al. 1996]{weisskopf96} Weisskopf, M. C., O'Dell,
S. L., \& van Speybroeck, L. P. 1996, Proc. SPIE, 2805, 2

\end{thebibliography}
\end{document}